# Deadline Aware Virtual Machine Scheduler for Grid and Cloud Computing

Omer Khalid
CERN
Geneva, Switzerland
Omer.Khalid@cern.ch

Ivo Maljevic
Soma Networks
Toronto, Canada
IMaljevic@somanetworks.com

Richard Anthony
University of Greenwich
London, United Kingdom
R.J.Anthony@gre.ac.uk

Miltos Petridis
University of Greenwich
London, United Kingdom
M.Petridis@gre.ac.uk

Kevin Parrott
University of Greenwich
London, United Kingdom
A.K.Parrott@gre.ac.uk

Markus Schulz
CERN
Geneva, Switzerland
Markus.Schulz@cern.ch

## ABSTRACT

Virtualization technology has enabled applications to be decoupled from the underlying hardware providing the benefits of portability, better control over execution environment and isolation. It has been widely adopted in scientific grids and commercial clouds. Since virtualization, despite its benefits incurs a performance penalty, which could be significant for systems dealing with uncertainty such as High Performance Computing (HPC) applications where jobs have tight deadlines and have dependencies on other jobs before they could run. The major obstacle lies in bridging the gap between performance requirements of a job and performance offered by the virtualization technology if the jobs were to be executed in virtual machines. In this paper, we present a novel approach to optimize job deadlines when run in virtual machines by developing a deadline-aware algorithm that responds to job execution delays in real time, and dynamically optimizes jobs to meet their deadline obligations. Our approaches borrowed concepts both from signal processing and statistical techniques, and their comparative performance results are presented later in the paper including the impact on utilization rate of the hardware resources.

## Categories and Subject Descriptors

D.2.7 [**Software Engineering**]: Distribution, Maintenance, and Enhancement – *portability*; Metrics – *performance measures*;

## 1. INTRODUCTION

In the past few years, Virtualization technology have remarkably shaped the way data centres have come about to increase their resource utilization by virtualizing more and more of computing applications to reap the benefits of lower support costs, higher mobility of the application, higher fault tolerance with lower capital cost [1, 4, 5]. This have led to the evolution of the idea of Grid computing [2] where resources were distributed geographically spanning over very many data centres and sharing resources among diverse community of users in decentralised fashion. But the rigid boundaries between very many systems in the "Grid" and the constraints they posed led to development of "Clouds" such as Amazon EC2 [25] where users could access computing and storage resources on-demand metered by per hour use. The terms Grid and Cloud are almost interchangeable since both aims to provide computing resources to their respective user communities through abstract and well defined set of interfaces. In this evolution over the last decade, virtualization have increasingly brought a paradigm shift where applications and services could be packaged as thin appliances and moved across distributed infrastructure with minimum disruption not only on the servers but across desktops in large organizations.

Despite these major developments, some fundamental questions remain unanswered especially for running HPC jobs in the virtual machines either deployed on the Grid or Cloud that how significant virtualization overhead could be under different workloads, and whether jobs with tight deadlines could meet their obligation if resource providers were to fully virtualizes their worker nodes [3].

Given this potential, we investigated how this technology could benefit ATLAS [6] (on of CERN's high-energy physics experiments) grid infrastructure and improve its efficiency by simulating its High Performance Computing (HPC) jobs on virtual machines.

This poses a particular challenge in scientific grids such as LCG[1] that have to serve the needs of diverse communities often with competing and opposite demands but it's simpler to manage in commercial clouds such as Amazon's Elastic Cloud Computing (EC2) [25] or scientific clouds like Nimbus [13] where user have clear understanding that they would be paying for per hour usage and their SLA would terminate when they stop to pay.

---

[1] Large Hadron Collider Computer Grid (LHC) and Open Science Grid (OSG).

This enables the users to estimate their workload estimates in a different manner as compared to HPC users where sometimes it's not possible to accurately estimate job execution times. This problem is further compounded when extended virtualization job duration results in increased *deadline miss rate*. We define *deadline miss rate* as a function of jobs not meeting their deadlines.

Our study attempts to provide a way forward to address the above mentioned challenges in a way which is transparent to the users without letting them know that their jobs are run in the virtual machine and tries to optimize the job execution rate.

## 2. MOTIVATION AND BACKGROUND

Since some jobs are more CPU or memory intensive than the others and vice versa, this requires a dynamic and intelligent resource scheduling which is adaptive as the nature of workloads at any given moment changes. By just throwing more resources to a virtual machine at the expense of other competing ones does not solves the research problem, which in fact leads to lower resource utilization. We aim to explore and investigate this area of research that how such a scheduling model could be achieved which not only maximizes the success rate of jobs while maintaining high resource utilization.

Our work is novel in a sense that it's extensible to translate our system parameters in to monetary terms for a scenario where a cloud provider attaches some currency value to cost and incentives in the system.

### 2.1 ATLAS Experiment

Since ATLAS experiment uses PanDA [7] software framework to submit jobs to the grid. In our previous experiments, we demonstrated how such an existing Grid application framework is modified to deploy grid jobs in virtual machines [10, 11] while delivering higher job performance by tweaking the parameters of Xen hypervisor [24].

Since various ATLAS jobs have different execution times ranging from 6hrs to 24hrs each, it wasn't feasible to run the actual jobs to quantify the performance of our virtual machine scheduling algorithm. Since we needed to run thousands of jobs to test the algorithm, to overcome this constraint we developed a simulator that resembles a typical compute node in a Grid or a Cloud.

## 3. Theoretical Model

We have *n* number of jobs $j_i$ from $\{1...n\}$ for slots *s* each with deadline $d_i$ and execution time $e_T$ where each $j_i$ requires $R_i$ amount of resource quantum (CPU, memory). The Service Level Agreement (SLA) for Atlas experiment guarantees one CPU core per $j_i$. Our scheduler is *truthful* as it's built on this assumption that the user is providing accurate resource requirements. It has also to be noted that in our experimental context, ATLAS experiment's jobs are of two types; user analysis jobs and production jobs. We focus on production jobs as they have higher priority since their output is used to calibrate the detectors, and their resource requirements are well known and tend to be truthful.

In our simulation model, the algorithm intelligently schedules the jobs and learns over time about the missed deadlines under various conditions and tries to predict whether $j_i$ would be meeting its deadline $d_i$, and if not then take appropriate measures to improve it chances in meeting $d_i$. We assume that $\forall j_i : e_T \leq d_i$. Since virtualization incurs a constant overhead, let it be coefficient $\partial$ and $e_V$ as virtual execution time than the new deadline would be $d_{new}$. This implies:

$$virtual\ exec.\ time = (duration * overhead) + duration \quad (1)$$

It can also be formally expressed as following:

$$e_V = e_T + \partial e_T, \{\partial \geq 0\} \Rightarrow d_{new} \approx e_V \quad (2)$$

*Unlimited deadlines:* For the purpose of proof, let *D* be the deadline for $j_i$. In this case, we show that if $D = \infty$ then every job will be meeting its deadline despite virtualization overhead thus it would be an *ideal scheduler*. This is the upper bound of our system which it would never be able to cross where is

$$\sum_{i=1}^{n} e_V \leq \sum_{i=1}^{n} D_i \quad (3)$$

*Tight deadlines:* In this case, we show that if each $j_i$ has $e_T = d_i$ and since $e_V = \partial e_T$, thus

$$\therefore \forall j_i : d_{new} > d_i \text{ and } d_{new} < e_V \text{ given } \Delta d \geq 0 \quad (4)$$

This would be the worst case (lower bound) scenario since all the jobs would be missing their deadlines, and if the batch system kills them all when they will exceed their individual allocated run time, then the system would be heavily underutilized since all these jobs have to be re-submitted and previously utilized resources would be considered as wasted.

Although in our experimental context, no monetary incentive is involved but to commercially schedule virtual machines, we could have introduced economics parameters in the system following the approach of Fledman et al for scheduling sponsored Google's advertisement slots to a set of bidders [9]. Since we abstract physical machine resources (CPU and memory) as resource units, the boundaries set for each resource slot *s* is between time interval $[0, t)$.

If *N* is the number of time units required for a job, then the $s_N$ is the number of slots a job needs to complete where *t* acts as frequency of the system to measure the slot booking and resource utilization ratio for a given time span.

$$s_N := \frac{e_V}{t} \quad (5)$$

If $t = \infty$, then the slots have unlimited life and it would not be possible to measure their utilization. But this dimension of scheduling is outside the scope of our deployment scenario since scientific grids doesn't involve monetary factors as an input for their set of scheduling parameters.

### 3.1.1 Performance Metrics

To access the performance of our scheduling algorithm, we define the following metrics in our systems:

- o System performance to measure total number jobs completed during a period of time.
- o Deadline miss rate representing the number of jobs missing their deadline, thus being terminated by the scheduler.
- o Utilization rate for the CPU and memory to measure how long each resource have been active

To allow the scheduling algorithm to respond to the system properties, we introduced duration-executed vs duration-remaining ratio, donated as $x$, for $j_i$ that is projected to miss its current deadline, and is determined by:

$$x_i = \frac{(job\ duration\ remaining - time\ to\ deadline)}{job\ duration\ remaining} \quad (6)$$

The first, and easiest method, is to set an acceptance threshold such that when $x_i < x\ threshold\ (X)$ jobs are accepted and rejected otherwise. The basic idea behind this approach is that it is expected that that acceptance of jobs beyond a certain threshold would be counter-productive as most of them would fail.

The adaptive threshold update structure we have adopted, shown in figure 1, has been motivated by similar structures that are used in communication systems and digital signal processing. Namely, there are many control loops that have a similar goal: to track a given reference value, such as time (time synchronizers), phase (phase locked loops), etc and most notably in adaptive gain control, where goal is to maintain a certain reference power level. A good overview of these techniques can be found in [8]

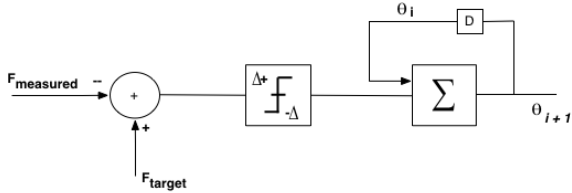

Fig 1. Adaptive $x$ threshold algorithm: $\theta_i$, $\theta_{i+1}$ - threshold values, D – delay element, $\Delta$ - delta step, $F_{target}$ – target failure rate, $F_{measured}$ – measured failure rate

In our implementation, a simple threshold update algorithm is based on trying to keep the failure rate close to a selected target failure rate. The threshold value for job acceptance is lowered if the failure rate increases and vice versa. The update step has been quantized to a small value $\Delta x$ in order to avoid fast threshold changes if the difference between the measured and targeted failure rate is large. The optimal value for the step $\Delta x$ has been determined through experimentation.

A third approach we have taken is to calculate the Probability Distribution Function (PDF) and Cumulative Distribution Function (CDF) of the success rate and use it to select the threshold value dynamically in a such way that $X_{Thresh}$ corresponds to the probability $P\ [x < X_{Thresh}] < P_{Thresh}$, where $P_{Thresh}$ is some pre-selected target probability of success. It is possible to use the CDF curve of the failure rate to drive the adaptive algorithm, but the number of successfully completed jobs is larger than the number of failed jobs, which results in more meaningful statistics. We have tested several $P_{Thresh}$ values to determine the best value. The CDF curve has been generated after the first 1000 jobs, and has been updated subsequently to reflect any possible changes in the system behavior.

Failure rate is representation of the jobs missing their deadlines in a certain window of time, independent of the global system performance, and derives the adaptive $x$ algorithm. Let donate $f$ as failure rate where as $F$ is the target failure for the system (0.2, 0.25]. There are two additional factors alpha $\alpha$ and success constant $S$ for $n$ number of jobs succeeding in meeting their deadline with total number of N jobs. Such method avoids having to keep track of an ever-increasing number of jobs, and rates in question are calculated as:

$$failure\ rate = 1 - ((n/N)(1 - \alpha) + S\alpha) \quad (7)$$

## 4. CASE STUDY

Since the present study is based on simulation results, we used the same machine resources configuration for the virtual server as of the physical servers with 4 CPU and 8GB RAM. For the training phase, to derive some core parameters we ran the simulator for job queue length of 10,000 hours of workload while for the steady phase the job queue length was 100,000 hours of workloads.

In our previous experiments, we focused only on the one set of job type that had the highest resource requirements with lower priority and proved through empirical results that para-virtualization, despite its over head, provide very neat solutions for the many of the problems faced by the Atlas Virtual Organization (VO). The questions we are trying to answer is:

- o How dynamic scheduling of workloads at the machine level will work once the batch system have scheduled a job to a particular node given that job will be executed in a virtual machine container?
- o What kind of scheduling technique could be used to optimize multiple virtual machines running HPC jobs?
- o Which parameter of the job could be considered as pivotal in scheduling policy; deadline-duration ratio or failure rate? Or both of them could be part of the same scheduling technique?
- o What would be the mix of executing job types on the machine to increase resource utilization?

All the above questions are of particular importance especially in the context of LHC's ATLAS experiment where short running *event generation* job have higher priority with low resource requirements over the competing long running *reconstruction* job having very high resource requirements. Thus, traditional scheduling methodologies could not be used if virtual machine based execution have to take place on the LCG Grid.

### 4.1 Training Phase

Since there are many different input parameters in the system such as resource ratio (memory to CPU), frequency of the scheduler, deadline buffer that was set to 5% of the job duration, alpha and delta values for the adaptive $x$ algorithm. We first ran the simulator in the training mode to establish optimum values for the above-mentioned parameters before running the actual simulation. In this section, results of this training phase are presented.

### 4.1.1 Resource Ratio Optimization

We ran training simulations for two different resource ratio, let it be donated as $R_i$ per job slot, configurations to measure the global success rate and the job deadline miss rate where resource (CPU: Memory) ratio were 1, 1.5, 2 and 3 (res_1, res_2, res_3, res_4) respectively. The number of jobs concurrently running in the systems is constraint by the available CPU since ATLAS [19] policy for Grid jobs is one job per CPU slot, so we kept the same policy for jobs being executed in the virtual machines.

Let $R_i = M/C$ where M donates memory and C donates number of CPU on the worker node.

Figure 2 shows the results for res_1 and represents the results for res_2. It has to be noticed that higher C reduced the overall system performance and increased the job deadline miss rate where as $R=2$ lead to the best performance, though jobs ran slower as it took more simulation time units, but more resources were available for the competing jobs. For all other experiments, we kept $R=1.5$ as golden middle resource ratio since it matched the configuration of physical servers.

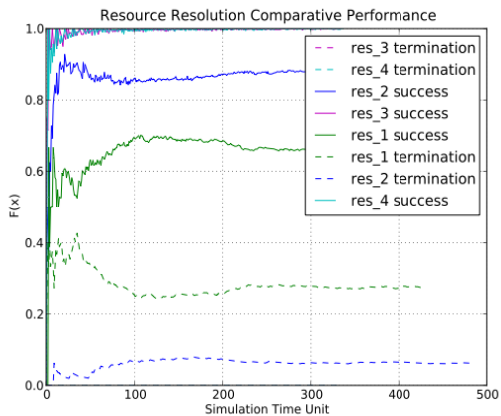

**Fig 2. Comparison of job success rate and termination rate for different resource configurations**

### 4.1.2 Alpha Delta Optimization

The next step was to train the algorithm for optimum alpha α and $\Delta x$ threshold values which could be then used for the *steady* phase. We ran the simulation with adaptive X scheduler and learning mode enabled for α [0.01, 0.05, 0.1] (configuration *alpha_1, alpha_2, alpha_3* respectively while keeping $\Delta x=0.1$) and $\Delta x$ [0.05, 0.1, 0.2] (*delta_1, delta_2, delta_3* respectively while keeping α $=0.01$). It was observed that as long as α and $\Delta x$ have a ratio factor between 5 -10, they performed better then rest of the combination of values as shown in the following table 1.

**Table 1. Global Success, Deadline Success rate and deadline miss rate for different alpha and delta values**

| Conf | Global success rate | Deadline success | Deadline miss |
|---|---|---|---|
| alpha_1 | 0.92 | 0.83 | 0.17 |
| alpha_2 | 0.93 | 0.86 | 0.14 |
| alpha_3 | 0.90 | 0.82 | 0.18 |
| delta_1 | 0.92 | 0.84 | 0.16 |
| delta_2 | 0.93 | 0.86 | 0.14 |
| delta_3 | 0.93 | 0.85 | 0.15 |

### 4.1.3 X threshold Optimization

The job distribution input dataset is randomized since *event generation, simulation* and *reconstruction* jobs have different durations and resource requirements, and system's deadline miss rate metric is driven by the virtualization overhead at a given moment which will affect whether a job will meet it's deadline or not when it first appears to miss it. This deadline ratio to job duration is expressed as *x* in the system. Although at any given moment when *x* is recorded to miss its deadline, it's success or failure is heavily influenced by the profile of the concurrent jobs which determines the virtualization over head as described in our previous study [11].

Since $x_i$ success or failure is linked with the evolution of virtualization overhead for the length of the job, so initial *x threshold* value for the system is significant since keep it too low will result in termination of jobs which otherwise might have succeeded and keeping it too high would have led to all jobs being allowed to run which might not eventually meet their deadlines, thus decreasing resource utilization.

We trained the algorithm for different *x* values [0.3, 0.7, +0.1] while keeping α and $\Delta x$ threshold as 0.05 and 0.1 respectively. Our results showed that the optimum *x* lied somewhere between [0.5,0.6] as shown in the figure 3 where failure rate converges to < 0.2 but with lower *x* threshold in the early phases leads to higher failure rate and then the algorithm responds to it by altering the *x* threshold. We selected *x*=0.6 as a *golden middle* for the initial threshold value for later experiments.

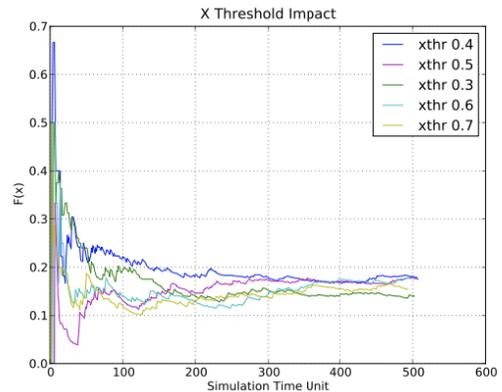

**Fig 3. Failure rate evolution for different x threshold values over the period of time**

## 4.2 Steady Phase

### 4.2.1 Configuration
Once the key optimization parameters were measured through the training phase, we ran the simulation for 100k hours of workloads for different set of configuration (alg_1, alg_2, alg_3, alg_4, alg_5) to measure that how HPC workloads will perform when ran on virtual machines. *Alg_1* is our physical baseline where as *alg_2* represents virtual baseline without the use of any optimization technique. Afterwards, we progressively tested various optimization techniques. *alg_3* employs *Virtual Dynamic* optimization where virtualization overhead was adapted without adaptive algorithm. *alg_4 (Virutal Dynamic Adaptive)* is a step-up function of *alg_3* using adaptive algorithm where as *alg_5 (Virtual Dynamic Statistical)* was ran using CDF to adapt the *x* threshold for the executing workloads.

### 4.2.2 Performance Results
Physical baseline had the best performance in overall system success rate with the minimum job deadline miss rate, as seen in figure 4, and since the core objective of this study was to develop an algorithm which could deliver the performance within 10%-15% range. Without any optimization technique, static VO (*alg_2*) lead to worst performance where deadline miss rate was 0.58 but the empirical data shows that consider performance gains could be made by alternating virtualization overhead (*alg_3*) according to the running workloads. Our adaptive algorithm (both *alg_4* and *alg_5*) further improved the job success rate from 0.78 to 0.84 by 7.7% while job deadline miss rate by ±0.17, which is 26%, less than *alg_3's* 0.23 deadlines miss rate.

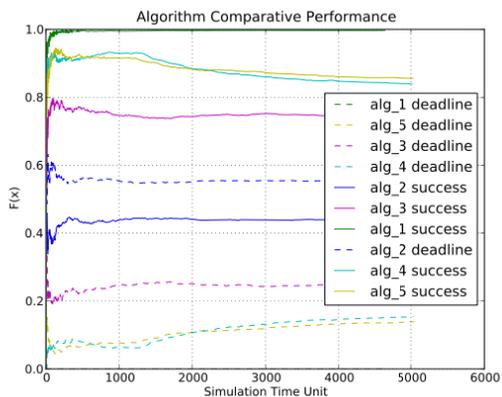

**Fig 4. Comparative performance of different algorithms tested for the simulation**

This study provided ample evidence that there is considerable room for improve in the performance of virtual HPC workloads by using dynamic VO which is driven by the CPU and memory characteristics of the workloads, and adapting *x* threshold at run time driven by the subset of near-past job history (success and failures). It has to be noted that we determined VO for ATLAS HPC workloads [11] and thus the VO levels we used have to be trained before adaptive *x* algorithm could be applied to workloads with different computational and memory requirements.

## 5. RELATED WORK
Present day virtual machine deployment engines such as OpenNebula [12], Nimbus [13] and others have focused on developing standard and transparent mechanisms and interfaces to deploy virtual machines on clusters managed which could be incorporated into scientific grids and clouds but does not address the issues of job performance or jobs meeting their deadlines which is critical for HPC experiments. Similarly, efforts [14, 15, 16] done to abstract compute resources and making them available as *utility computing* using the above-mentioned VM deployment engines but their research focus have been on developing leasing mechanisms for bringing up on-demand virtual clusters for parallel jobs.

The work done by VSched project [17] is very relevant to our research problem and complements our work as they have worked towards meeting job deadlines when batch and user-interactive (UI) applications are mixed and executed in virtual machines. Their underlying scheduling technique significantly different from our approach since user interactive applications require minimum response latency when deadline represents a response latency as compared to our use case where the issue at hand have been to optimize job success ratio for HPC batch jobs with no UI constraints. We addressed the issue through monitoring job success rates in real time where as Sadon et al [18] tackled the same problem by working towards a scheduling model which could reshape and re-size virtual machines as the work load changes under different conditions. Lingrand et al [19] have taken an alternative approach to apply optimization technique to job submissions and it would be very interesting to explore this further in the context of job submission to virtual machines.

Since a lots of efforts have been also put in to hotspot management and live migration of virtual machines as services demands for resources goes up and down during peak-offpeak times [20, 21] to consolidate VM's on few resources while powering down free resources to reduce electricity and cooling foot print. This requires that migrated VM's be migrated between servers with minimum interruption to their running applications by retaining network connectivity and checking point volatile memory [22, 23]. This is a very interesting arena of research but is beyond the scope of our study.

## 6. CONCLUSION
In this paper, we have presented a dynamic and real-time virtual machine scheduler that monitors job execution pathways and optimize job success rate for HPC workloads when ran on virtual machines in the scientific grids. To achieve this we developed a job execution simulator and implemented various scheduling techniques including statistical methods for prediction purposes. We observed that both statistical and adaptive *x threshold* models yielded best performance as compared to static virtualization model. Empirical data also proved that adaptation of virtualization overhead dynamical is critical for HPC jobs to deliver comparable performance when run in physical machines. We also learnt that resource resolution ratios affect more of job termination rate rather than job deadline rates, and for lower ratio job termination rate increased.

The limitation of our approach is that for a given set of workload types, initial performance metric has to be established to define virtualization overheads by running them in virtual machines before simulations could be run. Secondly, the algorithm also has to be trained for alpha and delta parameters for given workloads.

Our work is complementary to the ongoing developments taking place in the virtualization of grids and clouds where high level back clusters are increasingly providing virtualization interfaces, and could be extended in the scope of commercial cloud providers. We also wish to expand our future research from scheduling of virtual machines from a single node to cluster of nodes, and especially to investigate that how live migration of virtual machines could be integrated in such a manner that when a job is bound to miss it's deadline on this particular node, then rather than terminating it if it could be migrated to another resource in the cluster where it could meet its deadline obligation. This will further improve job throughput in the grid systems.

## 7. ACKNOWLEDGMENTS
We are grateful to ATLAS Computing and PanDA experiment and collaboration in providing support to us during our research.